\documentclass[aps,prb,twocolumn,showpacs,amsmath,amssymb,superscriptaddress]{revtex4-1}
\usepackage{graphicx}
\usepackage{dcolumn}
\usepackage{bm}
\usepackage{color}
\usepackage{graphicx}
\usepackage{epsfig}
\include{graphic}

\begin{document}


\title{Theory and experiment on cavity magnon polariton in the 1D configuration}

\author{B. M.~Yao} \email{yaobimu@mail.sitp.ac.cn}
\affiliation{Department of Physics and Astronomy, University of Manitoba, Winnipeg, Canada R3T 2N2}
\affiliation{National Laboratory for Infrared Physics, Chinese Academy of Sciences, Shanghai 200083, People$'$s Republic of China}

\author{Y. S. ~Gui}
\affiliation{Department of Physics and Astronomy, University of Manitoba, Winnipeg, Canada R3T 2N2}

\author{Y. ~Xiao}
\affiliation{College of Science, Nanjing University of Aeronautics and Astronautics, Nanjing 210016, China}
\affiliation{Department of Physics,  McGill University, Montr\'eal, Qu\'ebec, H3A 2T8, Canada}

\author{H. ~Guo}
\affiliation{Department of Physics,  McGill University, Montr\'eal, Qu\'ebec, H3A 2T8, Canada}

\author{X. S. ~Chen}
\affiliation{National Laboratory for Infrared Physics, Chinese Academy of Sciences, Shanghai 200083, People$'$s Republic of China}

\author{W. ~Lu}
\affiliation{National Laboratory for Infrared Physics, Chinese Academy of Sciences, Shanghai 200083, People$'$s Republic of China}

\author{ C. L. ~Chien}
\affiliation{Department of Physics and Astronomy, Johns Hopkins University, Baltimore, Maryland 21218, USA}

\author{C. -M. ~Hu}
\affiliation{Department of Physics and Astronomy, University of Manitoba, Winnipeg, Canada R3T 2N2}

\date{\today}

\begin{abstract}
We have theoretically and experimentally investigated the dispersion of the cavity-magnon-polariton (CMP) in a 1D configuration, created by inserting a low damping magnetic insulator into a high-quality 1D microwave cavity. By simplifying the full-wave simulation based on the transfer matrix approach in the long wavelength limit, an analytic approximation of the CMP dispersion has been obtained.  The resultant coupling strength of the CMP shows different dependence on the sample thickness as well as the permittivity of the sample, determined by the parity of the cavity modes. These scaling effects of the cavity and material parameters are confirmed by experimental data. Our work provide a detailed understanding of the 1D CMP, which could help to engineer coupled magnon-photon system.

\end{abstract}



\maketitle

\section{Introduction}
Coupling between electrodynamics and magnetization dynamics is a subject of long-standing interest \cite{Bloembergen1954, Lax1954, Mills1974, Sanders1974, Wende1976, Silva1999, Huebl2013, Tabuchi2014,Zhang2014, Goryachev2014, Bhoi2014, Haigh2015, Lambert2015, Abdurakhimov2015, Yao2015, Bai2015, Sandeep2015, Kostylev2015, Hu2015, Tabuchi2015, Zhang2015, Silva2015, Soykal2010, Cao2015}. At the resonant condition where the microwave frequency approaches the eigenfrequency of a magnon, two pieces of key physics stand out from the resonant magnon-photon coupling: (i) mode hybridization takes place which leads to the formation of a magnon polariton\cite{Mills1974}, (ii) damping correlation happens which causes radiation damping \cite{Bloembergen1954, Sanders1974, Wende1976, Bai2015, Silva2015}.

Very recently, interest in the physics of magnon-photon coupling has grown significantly due to the advancement of microwave cavity and spintronic techniques. Experimental progress has shown that the strong coupling between magnon and photon can be easily achieved by inserting a low damping magnetic material into a high quality microwave cavity\cite{Huebl2013, Tabuchi2014, Zhang2014, Goryachev2014, Bhoi2014, Haigh2015, Lambert2015, Abdurakhimov2015, Yao2015, Bai2015, Sandeep2015, Kostylev2015}. Such a cavity spintronic method \cite{Hu2015} creates new avenues for studying and utilizing both  spintronics and quantum information, as highlighted by the recent development of a cavity spin pumping method for coherently manipulating spin currents using magnon-photon coupling \cite{Bai2015} and the invention of quantum transducers that coherently link diverse quantum systems \cite{Tabuchi2015, Zhang2015}, respectively.

\begin{widetext}
\begin{table} [h]
\begin{center}
\label{table1} \caption{Summary of different theoretical approaches developed recently for studying the physics of magnon-photon coupling.}
\begin{tabular}{ccc}
  \hline\hline
  \\
  \textbf{Approaches} & \textbf{Classical theory} & \textbf{Quantum theory}\\
  \\
  \hline\hline
  \\
 \textbf{Intuitive}  & coupled harmonic oscillators [Bai \textit{et al.}, Ref. \onlinecite{Bai2015}]  & coupled harmonic oscillators [Zhang \textit{et al.}, Ref. \onlinecite{Zhang2014}]\\
  \\
  \hline
  \\
 \textbf{Concise}  & ~~ electrodynamic phase correlation [Bai \textit{et al.}, Ref. \onlinecite{Bai2015}] & ~ quantum entangled states [Soykal \textit{et al.}, Ref. \onlinecite{Soykal2010}]\\
  \\
  \hline
  \\
  ~  & ~ 1D scattering theory [Cao \textit{et al.}, Ref. \onlinecite{Cao2015}]~ & ~ \\
  \textbf{Detailed} & ~ & ~ \\
   ~        & ~ transfer matrix method [this work] & ~ \\
   \\
  \hline\hline

\end{tabular}
\end{center}
\end{table}
\end{widetext}

Theoretically, as summarized in the Table I, different approaches have been developed for studying the resonant magnon-photon coupling. The simplest approach roots on the physics of coupled harmonic oscillators, which provides an intuitive picture of the key physics of mode anticrossing and damping evolution. This approach can be formulated by using either a classical or a quantum mechanical model, as detailed in the Supplementary Materials of Ref. \onlinecite{Zhang2014} and \onlinecite{Bai2015}, respectively. Within such an approximation of harmonic oscillators, there is no distinction between the physics of quantum and classical coupling.

Such a distinction appears in the refined general approaches which specifically identify different origins of the magnon-photon coupling in the quantum and classical regimes. In the fully quantum-mechanical treatment\cite{Soykal2010}, it was found that the quantum-coherent magnet-photon coupling arose from the entangled quantum states of spin orientation and photon number. On the other hand, in a concise but general classical treatment independent of the sample and waveguide geometry \cite{Bai2015}, it was found that classical-coherent magnet-photon coupling appears due to the mutual electrodynamic coupling of the macroscopic microwave field and the macroscopic magnetization, which follows the classical phase correlations determined by Faraday's law and Amper\'{e}'s law (i.e. by Maxwell's equations). So far, it remains a theoretical challenge of clearly define the border between the quantum and the classical coupling regimes, which is needed for understanding the transition from quantum- to classical-coherent magnet-photon coupling. However, in light of the general classical model \cite{Bai2015} which quantitatively captures both key features of mode hybridization and radiation damping measured in recent experiments\cite{Huebl2013, Tabuchi2014, Zhang2014, Goryachev2014, Bhoi2014, Haigh2015, Lambert2015, Abdurakhimov2015, Yao2015, Bai2015, Sandeep2015}, it appears that many of these experiments probe the property of cavity magnon polaritons (CMPs) which are formed by the classical-coherent magnet-photon coupling.

As known from early theoretical study of the propagation of electromagnetic waves in a microwave waveguide partially filled by a magnetic slab\cite{Lax1954}, the details of the CMP tediously depend on the parameters of practical interest, such as the shape and size of both the waveguide and the sample. In general, it requires numerical solution of the coupled Maxwell's equations and Landau-Lifshitz-Gilbert equation. Only in special cases, simple solutions can be obtained. One of such special examples is the 1D cavity inserted with a magnetic slab, which was recently theoretically analyzed by using the 1D scattering theory\cite{Cao2015}. To verify the specific theoretical results obtained from such a detailed approach, an experimental method of studying the classical-coherent magnet-photon coupling in the 1D configuration needs to be developed.

In this paper, we develop such a 1D method experimentally. Furthermore, we simplify the 1D scattering theory by using the straightforward transfer matrix method, so that our experimental and theoretical method can be easily used in a combined way for studying the detailed physics of the CMP in the 1D configuration. One of our focus here is to explain how the coupling between photons and magnons depends on the thickness of the sample as well as the parity of cavity mode for a 1D cavity.

This paper is split into two main sections: theoretical model and experimental results. In the theoretical model section, we first provide a brief description of the 1D cavity used in this work and use the transfer matrix approach to derive the cavity resonance. Then the dispersion of the CMP due to the hybridization of the cavity resonance and the ferromagnetic resonance (FMR) in a bulk yttrium iron garnet (YIG, Y$_3$Fe$_5$O$_{12}$) is derived by solving the coupled Maxwell's equations and Landau- Lifshitz-Gilbert equation using the transfer matrix method. To highlight the coupling strength in a CMP, the full-wave simulation is simplified in the limit of long wavelength, which clearly shows its strong dependence on the cavity length, sample thickness, sample permittivity as well as the parity of the cavity modes. Finally we present experimental results to confirm our numerical results and analytical model.

\section{Theory of cavity magnon polariton in the 1D configuration}

For a quantitative understanding of the CMP dispersion, we developed a model to solve the case for a 1D microwave cavity. In general it is difficult to provide the exact analytical solution of modes excited in a ferrite loaded cavity under a magnetic bias, where the propagating wave encounters spin dynamics in the loaded ferrite sample\cite{Lax1954,textbookferrite_Soohoo, textbookferrite_Lax}. However, in the special case of the one dimensional cavity that is used in this work the problem of wave propagation can be analytically solved.\cite{Yao2015} To simplify the complicated mathematical process, the scattering matrix treatment and transfer matrix, being derived from general principles such as conversion of energy, are used to deduce the exact solution of the CMP. This full-wave solution provides all detailed information of the coupled magnon-photon system without any adjustable parameters. In addition, a simplified analytical solution is also deduced in the long wavelength limit and validated by the full-wave calculation, allowing the prediction of the coupling strength between cavity modes and magnon modes in a straightforward way. 

\subsection{Microwave cavity based on waveguide assembly}

\begin{figure} [t]
\begin{center}
\epsfig{file=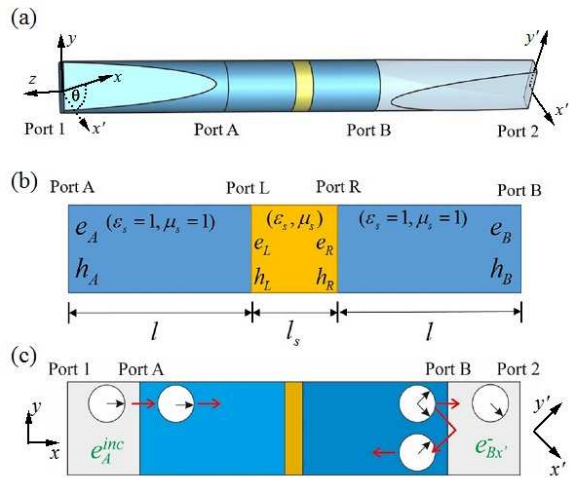,width=7.8cm}
\caption{(Color online) (a) Schematic of the waveguide assembly cavity. Two circular-rectangular transitions are twisted by an angle $\theta$ with respect to each other and are connected to a vector network analyzer through SMA to rectangular waveguide adapters. (b) Wave propagation between port A and B in (a) described by a cascade connection of two-port networks. (c) Schematic diagram of the wave propagation for incident wave from port 1 and  transmitting wave toward port 2. For the coupling experiments, the static magnetic bias $H_{ext}$ is applied in the $x-y$ plane, perpendicular to the direction of wave propagation (along $z$ direction). }\label{Fig1}
\end{center}
\end{figure}

As shown in Fig. \ref{Fig1}(a) the microwave cavity used in this work is a Fabry-Perot-like resonator based on the Ku-band (12 GHz- 18 GHz) assembled waveguide apparatus\cite{Carot2012, Sounas2013, Yao2015}, where circular waveguides are connected through circular-rectangular transitions to coaxial-rectangular adapters and two transitions are rotated by an angle $\theta$. The microwave propagation in such an apparatus can be characterized by the S-parameter and measured by a vector network anlayzer.

The fundamental mode of the rectangular waveguide is TE10 and the polarization of the microwave electric field is along $x$ and $x^\prime$ (the short axis of the rectangle in Fig. \ref{Fig1}(a)) for port A and B, respectively. For the circular waveguide the fundamental mode is TE11, supporting two degenerate orthogonal waves, named $x-$ and $y-$polarized waves for port A ($x^\prime-$ and $y^\prime-$polarized waves for port B). Therefore, a circular-rectangular transition is designed to smoothly transform the TE10 mode of its rectangular port (port 1) to the co-polarized TE11 mode of its circular port (port A) without any reflection. However, propagating from port A towards port 1 only $x$-polarized waves be can transmitted through while $y-$ polarized waves are reflected with a reflection coefficient of $R$ (very close to 1) and a phase shift of $\phi_{y}$, which can be determined experimentally. Similar effects appear for $x^\prime$- and $y^\prime-$ polarized waves propagating form port B toward port 2. Because of this mirror-like effect for $y-$ and $y^\prime-$ polarized waves, a Fabry-Perot-liked resonator is built and a standing wave is generated in the long dimension (along the $z$ direction) of the circular waveguide at a particular frequency, which can be deduced by Maxwell's equations as detailed in the discussion below.

\subsection{Analytical model of the resonant cavity}
\label{Section:Analytical model of the resonant cavity}

To provide the analytical solution for the wave propagation in the microwave cavity and hence the quantitative analysis of the coupling between magnon and cavity modes, we start with the transfer matrix (the ABCD matrix in microwave terminology) for a dielectric loaded cavity. As shown in Fig. \ref{Fig1}(b), $l_s$ is the thickness of the sample loaded waveguide and $l$ is the length of air-filled waveguide. Hence the whole length of the circular waveguide is $W$=2$l$+$l_s$. Following the $e^{-i\omega{t}}$ convention, the transfer matrix for a two-port network, for example, the sample-loaded waveguide (yellow area between port L and R in Fig.\ref{Fig1}(b)), can be defined as \cite{Pozar2012}
\begin{equation}
\left(\begin{matrix} e_L\\ h_L \end{matrix}\right)=\left(\begin{matrix}
\cos({k\!_s}{l\!_s}) & -iZ_s\sin({k\!_s}{l\!_s})\\
-iZ_s^{-1}\sin({k\!_s}{l\!_s}) & \cos({k\!_s}{l\!_s})
\end{matrix}\right)\left(\begin{matrix} e_R\\ h_R \end{matrix}\right)=M_s\left(\begin{matrix} e_R\\ h_R \end{matrix}\right)
\label{Eq:M_matrix}
\end{equation}

\noindent where the subscript $L$($R$) indicates the left(right) interface of the sample. Equation (\ref{Eq:M_matrix}) links the electric field ($e$) and magnetic field ($h$) of microwave propagation in a dielectric medium using the ABCD matrix $M_s$, where the matrix elements relate only on the material parameters of wave number ($k\!_s=(\omega/c)\sqrt{\varepsilon\!_s\mu _s-(\omega\!_c/\omega)^2}$), the characteristic impedance ($Z\!_s = \omega {\mu_0}{\mu\!_s}/k\!_s$) and the length of transmission line ($l\!_s$). Here $\omega$, $\omega \!_c$, $\varepsilon_s$($\mu_s$), and $\varepsilon_0$($\mu_0$) are the frequency of the operating microwave, the cut-off frequency of the circular waveguide, the relative complex permittivity (permeability), and vacuum permittivity(permeability), respectively. $c=1/\sqrt{\varepsilon_0\mu_0}$ is the vacuum speed of light. For the air-filled region in waveguide, the wave number and characteristic impedance are $k_0$ and $Z_0$ using $\varepsilon\!_s=\mu\!_s=1$. 

We note that the finite lateral size of the waveguide results in the cut-off wavelength $\lambda_{c}=2\pi{c}/\omega\!_c$; for the TE11 mode in a circular waveguide  $\lambda_c$ is 3.41 times the radius (8.05 mm for standard Ku-band circular waveguides)\cite{Pozar2012}, corresponding to $\omega\!_c/2\pi=10.85$ GHz. As a consequence, the velocity of microwave propagation along the waveguide is less than its velocity through free space (speed of light), which results in a reduction of the ratio of the effective sample thickness with respect to the effective length of the cavity and hence significantly affects the dispersion of the CMP.

The convenience of the transfer matrix relies on the fact that the ABCD matrix of the cascade connection of the two networks is equal to the product of the ABCD matrices representing the individual two-ports. As such in Fig. \ref{Fig1}(b), the transfer matrix between ports A and B is $M_{AB}=M_1M_sM_2$ and $M_1$ and $M_2$ are the air-filled circular waveguide located at the left and right side of the sample, respectively. However, the ABCD transfer matrix relates the total electric and magnetic field at the ports;  to characterize the fields incident on the ports to those reflected from the ports the scattering matrix (also referred to as $S$-parameter) should be involved, defined as
\begin{equation}
\left(\begin{matrix}e_{A}^{-} \\ e_{B}^{-} \end{matrix}\right)=
\left(\begin{matrix} S_{AA} &S_{AB}\\ S_{BA} &S_{BB}\end{matrix}\right)
\left(\begin{matrix} e_{A}^{+} \\e_{B}^{+}\end{matrix}\right),
\label{Eq:SAB}
\end{equation}

\noindent where the superscript \textquotedblleft$+$\textquotedblright/\textquotedblleft$-$\textquotedblright indicate the waves entering/exiting the port. The elements of the scattering matrix can be easily obtained from the ABCD transfer matrix according to Eq. (\ref{SEq:SAB_matrix}) in Appendix \ref{Sec:appendixA}.\cite{Pozar2012} The great advantage of using the scattering matrix is that the S-parameter can be measured directly with a vector network anlayzer.

As the experimentally accessible parameter is the transmission coefficient $S_{21}$ between ports 1 and 2 of the assembled waveguide cavity, a wave entering or exiting the ports A or B must be considered by including the boundary condition. A detailed mathematical process can be found in Appendix \ref{Sec:appendixA}, leading to the transmission coefficient $S_{21}$ from port 1 to port 2
\begin{widetext}
\begin{equation}
S_{21}=\frac{{\cos (\theta ){S_{AB}}[ - 1 + {R^2}{e^{2i{\phi _y}}}{S_{AB}}{S_{BA}} + {e^{i{\phi _y}}}R{S_{BB}} - {e^{i{\phi _y}}}R{S_{AA}}( - 1 + {e^{i{\phi _y}}}R{S_{BB}})]}}{{ - 1 + {e^{2i{\phi _y}}}{{\cos }^2}(\theta ){R^2}{S_{AB}}{S_{BA}} + {e^{i{\phi _y}}}R{S_{BB}} - {e^{i{\phi _y}}}R{S_{AA}}( - 1 + {e^{i{\phi _y}}}R{S_{BB}})}}
 \label{Eq:S21}
\end{equation}
\end{widetext}

This equation is the key result of the full-wave solution for calculating the property of CMP in the 1D wave propagation configuration. Here $R$ and $\phi_{y}$ are the reflection coefficient and phase shift for the non-supported polarized microwave for the rectangular waveguide, respectively. For an air-filled waveguide (empty waveguide, $\varepsilon\!_s=\mu\!_s=1$), it is found that $S_{AA}=S_{BB}=0$ and $S_{AB}=S_{BA}=e^{ik_0(2l+l_s)}$ and hence
\begin{equation}
S_{21}=\cos(\theta)e^{ik_0(2l+l\!_s)}\dfrac{1-R^2e^{2i(2k_0l+k_0l\!_s+\phi_{y}})}{1-R^2\cos^2(\theta)e^{2i(2k_0l+k_0l\!_s+\phi_{y})}}.
\label{Eq:S21 emptycavity}
\end{equation}

The waveguide assembly acts as a transmission through waveguide with $S_{21}=e^{ik_0(2l+l_s)}$ at $\cos(\theta)=1$, while it would block transmission with $S_{21}=0$ at $\cos(\theta)=0$. Except for these extreme cases, the spectra of $S_{21}$ appears as a set of cavity modes occurring at a cavity resonant frequency of $\omega_{C\!R}$ with a minimum transmission of $|S_{21}(\omega_{C\!R})|=\cos(\theta)(1-R^2)/[1-R^2\cos^2(\theta)]\approx0$, where $k_{C\!R}(2l+l\!_s)+\phi_{y}=q\pi$ with $k_{C\!R}=(\omega_{C\!R}/c)\sqrt{1-(\omega_c/\omega_{C\!R})^2}$, $q$ being an integer.

\begin{figure} [b]
\epsfig{file=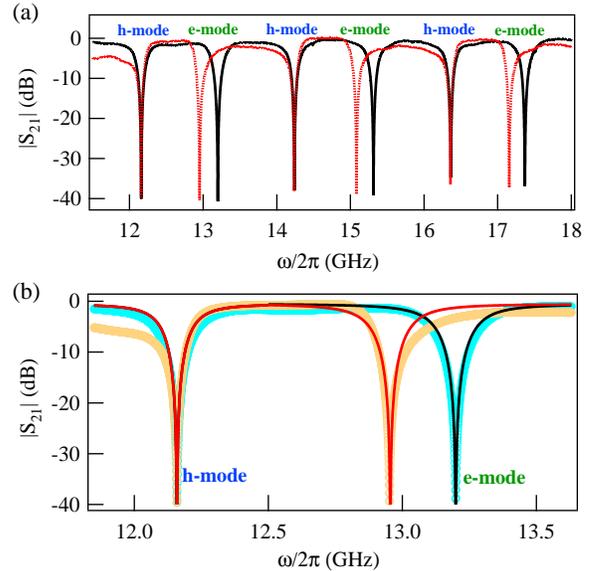,width=8cm}
\caption{(Color online) (a) Measured amplitude of $S_{21}$ spectra for the air-filled cavity (solid line) and Teflon-loaded cavity (dotted line) at $\theta=45^\circ$ and $H_{ext}=0$ with $W$=85 mm and $l_s$=1.6 mm.(b) Calculated $|S_{21}|$ spectra according to Eq. (\ref{Eq:e-h mode}) for air-filled cavity (Black line) and Teflon-loaded cavity (red line) as compared to the measured data (symbols). }
\label{Fig2}
\end{figure}

In our experiments, the angle of $\theta$ is rotated to be 45$^\circ$ unless otherwise specified. As expected a set of resonances is observed for the air-filled waveguide as well as the Teflon-loaded waveguide shown in Fig. \ref{Fig2}(a). These cavity resonances can be classified into two groups: \textbf{e-modes} (corresponding to even number $q$) being anti-nodes of the microwave electric field  and \textbf{h-modes} (corresponding to odd number $q$) being nodes of the microwave electric field at the sample position. In the absence of any external applied static magnetic bias, i.e., $H_{ext}=0$,  the loaded dielectric sample only causes the red-shift of the $e$-modes but has negligible effect for the $h$-modes; this effect can be clearly seen in Fig. \ref{Fig2}(a) when a 1.6 mm thick Teflon disk is loaded in the cavity. Using the full-wave simulation base on Eq. (\ref{Eq:S21}), the precise values of $R$ and $\phi_{y}$ were determined for the air-filled waveguide: for example, $R$=0.995 and $\phi_{y}=-36.1^\circ$ for the $e$-mode occurred at 13.199 GHz and $R$=0.995 and $\phi_{y}=-6.5^\circ$ for the $h$-mode occurred at 12.159 GHz. 

In order to distinguish the different dielectric dependence of $e$- and $h$-modes on the loaded material, we have simplified Eq. (\ref{Eq:S21}) near the cavity modes by Taylor expansion. In the limit of long wavelength, to $|k\!_sl\!_s|\ll{1}$, and assuming $R=1$ for simplicity, one can obtain the first order approximation given by
\begin{equation}
\begin{matrix}
S_{21}\!=\!\dfrac{2\sqrt{2}}{3}e^{-i\phi_y}\!\left[ 1 - \dfrac{1}{1-i[6l(k_0-k_{C\!R})+3l\!_s(\frac{k\!_s^2}{\mu\!_sk_0}-k_0)]}\right], \vspace{1mm}\\
\hspace{6cm} \mbox{(\textit{e}-mode)} \vspace{1mm}\\
S_{21}\!=\!-\dfrac{2\sqrt{2}}{3}e^{-i\phi_y}\!\left[ 1 - \dfrac{1}{1-i[6l(k_0-k_{C\!R})+3l\!_s(\mu\!_s-1)k_0]}\right]. \vspace{1mm}\\  
\hspace{6cm}\mbox{(\textit{h}-mode)} \vspace{1mm}
\end{matrix}
\label{Eq:e-h ks mode}
\end{equation}

The prefactor of $S_{21}$ is determined to be $2\sqrt{2}/3$ since far from the cavity modes $|S_{21}|$ tends to be $2|\cos(\theta)|/[1+\cos^2(\theta)]$ according to Eq. (\ref{Eq:S21 emptycavity}). To clearly see the resonant feature of the cavity modes, we rewrote the above expressions as
\begin{equation}
\begin{matrix}
S_{21}=\dfrac{2\sqrt{2}}{3}e^{-i\phi_y}\left[ 1 + \dfrac{\Delta\omega_{C\!R}}{i(\omega-\omega_{C\!R}+\delta\omega_{C\!R})-\Delta\omega_{C\!R}}\right],  \vspace{1mm}\\
\hspace{6cm}\mbox{(\textit{e}-mode)} \vspace{1mm}\\
S_{21}=-\dfrac{2\sqrt{2}}{3}e^{-i\phi_y}\left[ 1 +\dfrac{\Delta\omega_{C\!R}}{i(\omega-\omega_{C\!R}+\delta\omega_{C\!R})-\Delta\omega_{C\!R}}\right], \vspace{1mm} \\
\hspace{6cm} \mbox{(\textit{h}-mode)} \vspace{1mm}
\end{matrix}
\label{Eq:e-h mode}
\end{equation}

with
\begin{equation}
\begin{matrix}
\delta\omega_{C\!R}&=&\dfrac{l_s}{2l}\omega\left[(\varepsilon\!_s-1)+\left(1-\dfrac{1}{\mu\!_s}\right)\left(\dfrac{\omega\!_c}{\omega_{C\!R}}\right)^2\right], \mbox{(\textit{e}-mode)}\\
\delta\omega_{C\!R}&=&\dfrac{l_s}{2l}\omega\left(\mu\!_s-1\right)\left[1-\left(\dfrac{\omega\!_c}{\omega_{C\!R}}\right)^2\right].  \hspace{1.5cm} \mbox{(\textit{h}-mode)}
\end{matrix}
\label{Eq:e-h mode shift}
\end{equation}

Here the linewidth of the cavity mode is $\Delta\omega_{C\!R}=(c/6l)\sqrt{1-(\omega\!_c/\omega_{C\!R})^2}$. Physically, the linewidth of the resonator is related to the energy loss of the cavity loaded in the circuit, which can be estimated by the lifetime of the photon trapped in the cavity. Due to the finite cross section of waveguide, the velocity of the microwave photon at resonances is $c\sqrt{1-(\omega\!_c/\omega_{C\!R})^2}$ along the waveguide, and hence the travel time is about $\tau_t=2l/c\sqrt{1-(\omega_c/\omega_{C\!R})^2}$. Depending on the polarization, half of the photons could be lost at the interface of either port A or B in our proposed 1D cavity; thus, the average lifetime of photon is about $\tau_l=\tau_t\sum\limits_{n = 1}^{ + \infty }n2^{-n}=2\tau_t$, which is the same order as  $1/\Delta\omega_{C\!R}=3\tau_t$ estimated from the transfer matrix approach.

Equations (\ref{Eq:e-h mode}) and (\ref{Eq:e-h mode shift}) clearly indicate that the $h$-mode is not sensitive to the dielectric material (usually $\mu\!_s=1$) loaded at the center of the cavity where the node of the microwave electric field is present. 
Using the values of  $\varepsilon_s=2.0$ (at 13 GHz provided by the company), $\omega_{C\!R}/2\pi$=13.199 GHz (12.159 GHz) for the $e$-mode ($h$-mode), and $l\!_s=1.6$ mm, the measured $S$-parameter for the dielectric loaded cavity can be well reproduced as shown in Fig. \ref{Fig2}(b) based on the calculation according to Eq. (\ref{Eq:e-h mode}). The red-shift of the $e$-mode shows excellent agreement between measurement and calculation with $\delta\omega_{C\!R}/2\pi$=0.25 GHz.  The calculated linewidth of $\Delta\omega_{C\!R}/2\pi=(c/6l)\sqrt{1-(\omega\!_c/\omega_{C\!R})^2}=0.11$ GHz is slightly smaller than the measured linewidth of $\Delta\omega_{C\!R}/2\pi=0.14$ GHz, which may be due to the contribution from intrinsic damping caused by factors such as conductor loss of the cavity, not included in our model. The intrinsic quality factor $Q_{0}$\cite{Kaiser2010,Russer} of cavity mode which denotes the loss in the resonator itself can be estimated as $\sim$ $10^{3}$ approximately according to the ratio between cavity resonance frequency and the bandwidth at +3dB with respect to the minimal transmission \cite{Khanna1983}.

\subsection{Full-wave solution of the CMP dispersion}

Far from the FMR frequency $\omega_{F\!M\!R}$ of the ferrite sample, the spin-precession frequency is either much larger or much smaller than the frequency ($\omega$) of the driving microwave field. Therefore, the interaction between the electromagnetic field and spin dynamics is negligible. As a consequence, the YIG sample acts as a dielectric medium, only causing the red-shift of the $e$-mode while causing no first order effect for the $h$-mode as discussed in Sec. \ref{Section:Analytical model of the resonant cavity}. In the vicinity of the FMR frequency, the dynamic response of the spin precession $\mathbf{m}$ driven by the microwave magnetic field $\mathbf{h}$ is governed by the Landau-Lifshitz-Gilbert equation\cite{Gilbert2004}. Meanwhile, the wave propagation in the ferrite should also satisfy Maxwell's equation. 

When $\omega_{F\!M\!R}$ approaches $\omega_{C\!R}$, the electrodynamics and magnetization dynamics are strongly coupled. Conventionally, one can solve the coupled Landau-Lifshitz-Gilbert equation and Maxwell's equation as detailed in the discussion in Appendix \ref{Sec:appendixB} by introducing the effective permeability $\mu_{e\!f\!f}$ given by
\begin{equation}
\mu_{e\!f\!f}=\dfrac{(1+\chi_L)^2-\chi_T^2}{1+\chi_L},
\label{Eq:mu_eff}
\end{equation}

\noindent where $\chi_L$ and $\chi_T$ are the diagonal and non-diagonal elements of the Polder tensor as defined in Eq. (B5) in Appendix \ref{Sec:appendixB}. The static magnetic bias $H_{ext}$ is applied along the $y^\prime$-direction, perpendicular to the direction of wave propagation (along $z$ direction) in a coordinate shown in Fig. \ref{Fig1}.

When $|\omega/\omega_{F\!M\!R}-1|\ll{}1$, the resonant feature of $\mu_{e\!f\!f}$ can be expressed as
\begin{equation}
\mu_{e\!f\!f}-1\approx- \frac{1}{2}\frac{(\omega _H +\omega _M)\omega_M/\omega}{(\omega-\omega_{F\!M\!R})+i\alpha (\omega _H + \omega _M/2)},
\label{Eq:resonant mu_s}
\end{equation}
\noindent with $\omega_H=\gamma{H_{ext}}$, $\omega_M=\gamma{M_0}$ and $\omega_{F\!M\!R}=\gamma\sqrt{H_{ext}(H_{ext}+M_0)}$. Here $\gamma=\mu_0e/m_e$ is the effective gyromagnetic ratio of an electron (charge $e$ and mass $m_e$), $\alpha$ is the Gilbert damping parameter, and $M_0$ is the saturation magnetization.

\begin{figure}[t]
\epsfig{file=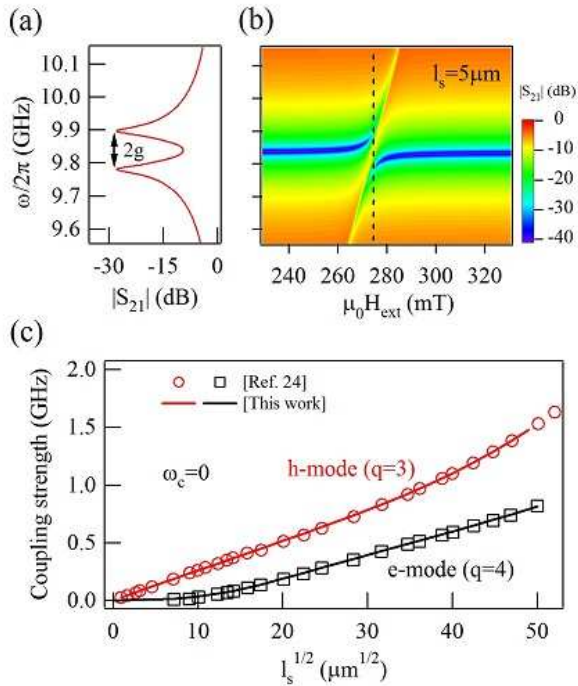,width=8cm}
\caption{(Color online) (a) $S_{21}$ spectra at $\omega_{F\!M\!R}=\omega_{C\!R}$=9.77 GHz, where the CMP gap was used to determine the coupling strength $g$. (b) Dispersion of the CMP as a function of microwave frequency and magnetic field at $l_s$=5 $\mu{m}$. The dashed line indicates the condition of $\omega_{F\!M\!R}=\omega_{C\!R}$. (c) Coupling strength with $q$=3 and 4 are calculated based on the scattering theory (open symbols, from Ref. \onlinecite{Cao2015}) and transfer matrix theory (solid lines, this work) with parameters of $\omega\!_c$=0, $\mu_0M_0=175$ mT, $\gamma=$ 176 $\mu_0$ GHz/T and $W=46$ mm.}
\label{Fig3}
\end{figure}

Replacing $\mu_s$ in Eq. (\ref{Eq:appendix kandZ}) with the effective permeability $\mu_{e\!f\!f}$, and substituting the resultant wave number $k_s$ and the characteristic impedance $Z_s$ for the YIG sample into the transfer matrix, the S-parameters ($S_{AA}$, $S_{AB}$, $S_{BA}$ and $S_{BB}$) between port A and B are deduced as detailed in the discussion in Sec. \ref{Sec:appendixA}. Then the transmission coefficient near the FMR can be calculated by the full-wave solution according to Eq. (\ref{Eq:S21}). With parameters of $\omega\!_c$=0, $\phi_y$=0, $R$=1, $\mu_0M_0=175$ mT, $\gamma=$ 176 $\mu_0$ GHz/T, $\alpha=1.25\times10^{-3}$, and $W=46$ mm, Figure \ref{Fig3}(b) plots the calculated $S_{21}$ spectra as a function of microwave frequency and magnetic field near the $h$-mode ($q$=3) occurred at $\omega/2\pi=$9.77 GHz. Without going into the detailed behaviour of the CMP, we focus on the coupling strength ($g$) of the CMP, which is determined by the CMP gap occurring at $\omega_{F\!M\!R}=\omega_{C\!R}$ as shown in Fig. \ref{Fig3}(a), to highlight the general feature of the coupling between the magnon and the photon in a 1D cavity. 

Under such an approximation of zero cut-off frequency ($\omega_c=0$), we compare the results from different approaches: the scattering theory\cite{Cao2015} and transfer matrix theory. The calculated cavity modes corresponding to $q=3$ and $q=4$ for the empty cavity occur at 9.77 GHz ($h$-mode) and 13.03 GHz($e$-mode), respectively, which are very close to the results obtained from the scattering theory. The coupling strength (solid lines) for both $h$-mode and $e$-mode are plotted as a function of $l_s^{1\!/2}$ shown in Fig. \ref{Fig3}(c). Despite significant differences in the microwave feed for the cavity used in the scattering theory (open symbols) and transfer matrix theory (solid lines) to model the CMP, the final result of the coupling strength is consistent, indicating the universality of the CMP, which is independent of the specific design of the 1D cavity.  

\begin{figure}[t]
\epsfig{file=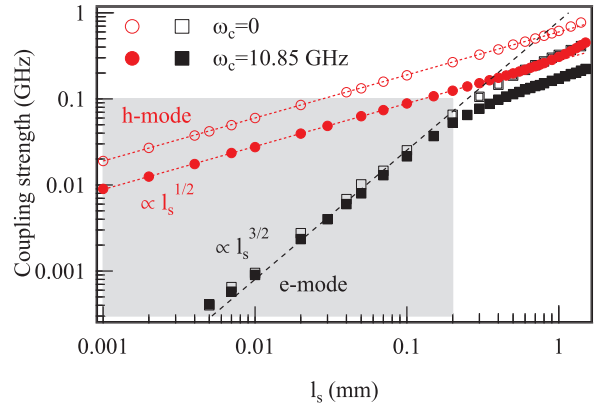,width=8cm}
\caption{(Color online) The coupling strength for neighbouring $h$-mode (circles) and $e$-mode (squares) in a 1D cavity (with $W$=85 mm) for $\omega_c$=0 (open symbols) and $\omega_c$=10.85 GHz (solid symbols).  For $\omega_c$=0, $h$-mode ($e$-mode) occurs at 12.34 GHz(14.11 GHz) at $\phi_y=0$ and $R=0.995$, while for $\omega_c$=10.85 GHz, $h$-mode ($e$-mode) occurs at 12.28 GHz(13.20 GHz) at $\phi_y=-36.5^\circ$ and $R=0.995$.  The dashed lines are guided by $\propto{}l_s^{1\!/2}$  and $\propto{}l_s^{3\!/2}$ relations for the $h$- and $e$-modes, respectively, in the limit of the long wavelength approximation. }
\label{Fig4}
\end{figure}

In a practical implementation, the impact of the cut-off frequency should be taken into account because of the finite lateral size of the 1D cavity. To demonstrate this effect, the coupling strength were calculated at $\omega\!_c$=0 and $\omega\!_c$=10.85 GHz (determined by the size of the circular waveguide in our cavity) by varying $l\!_s$ more than three orders of magnitude from 1 $\mu$m to 1.5 mm. For the $h$-mode, the coupling strength has about a factor of 2 reduction in the entire calculated $l\!_s$ range when using $\omega_c$=10.85 GHz as shown in Fig. \ref{Fig4}. In contrast, the coupling strength for $e$-mode is not sensitive to $\omega\!_c$ in the limit of long wavelength $l\!_s<0.2$ mm (indicated by the grey area in Fig. \ref{Fig4}), in which $(\sqrt{\varepsilon_s}\omega_{C\!R}/c)l\!_s<0.2$ with $\varepsilon\!_s\approx$15 for the YIG sample\cite{YIG1, YIG2}. 

Figure \ref{Fig4} also shows that the coupling strength for the $h$-mode and $e$-mode follows different power dependence of $l_s$ guided by the dotted lines in the limit of long wavelength. Beyond this range the red-shift of the $e$-mode causes its approach to the neighboring $h$-mode and the hybridization of $e$- mode and $h$-mode due to the dielectric coupling in the cavity may complicate the determination of the intrinsic coupling strength of the magnon-photon system. In addition, the cavity damping parameters $\Delta\omega_{C\!R}$ of both $h$-mode and $e$-mode are strongly dependent on $\omega_c$ (not shown). The quantification of cavity damping is very important to characterize the phenomena associated with distinct coupling range for the coupled magnon-photon system.

\subsection{Analytic solution of the coupling strength}
\label{Sec:Analytic solution of the coupling strength}

To verify the different scaling effects for $h$- and $e$- modes and hence provide an analytical solution of the coupling strength between magnons and photons in a 1D cavity, we have simplified the dispersion of the CMP in this section. To do so, Equation (\ref{Eq:resonant mu_s}) is substituted into Eq. (\ref{Eq:S21}) and the Taylor expansion of $l\!_s$ is used to determine the dominant power dependence of the coupling strength.

\subsubsection{Coupling strength $g$ for $h$-modes}

In the limit of long wavelength, the first order approximation is sufficient to model the dispersion of the CMP associated with $h$-modes. In this case, the transmission coefficient near the FMR is reduced to
\begin{eqnarray}
S_{21}&=&-\frac{2\sqrt{2}}{3}e^{-i\phi_y} \nonumber \\
&&\times\left[ 1 + \frac{\Delta\omega_{C\!R}}{i(\omega-\omega_{C\!R})-\Delta\omega_{C\!R}+\frac{g^2}{i(\omega-\omega_{F\!M\!R})-\Delta\omega_{F\!M\!R}}}\right] \nonumber\\
\label{Eq:coupled h-mode}
\end{eqnarray}

\noindent with
\begin{eqnarray}
&&\Delta\omega_{C\!R}=(c/6l)\sqrt{1-(\omega_c/\omega_{C\!R})^2},\nonumber \\
&&\Delta\omega_{F\!M\!R}=\alpha(\omega _H + \omega _M/2),\nonumber \\
&&g^2=(l\!_s/4l)[1-(\omega_c/\omega_{C\!R})^2](\omega _H +\omega _M)\omega_M,
\label{Eq:homdecouplingstrength}
\end{eqnarray}

\noindent where $\Delta\omega_{C\!R}$ is the damping coefficient for the cavity resonance, $\Delta\omega_{F\!M\!R}$ is the damping coefficient for FMR and $g$ is the coupling strength between them. In Eq. (\ref{Eq:homdecouplingstrength}), the coupling strength $g$ for $h$-mode is linearly proportional to $l_s^{1\!/2}$ , $l^{-1\!/2}$ and independent of the permittivity $\varepsilon\!_s$ of the sample. It was also found that the cut-off frequency $\omega_c$ causes significant reduction of the coupling strength $g$ of the $h$-mode as well the damping of the cavity.

\begin{figure}
\epsfig{file=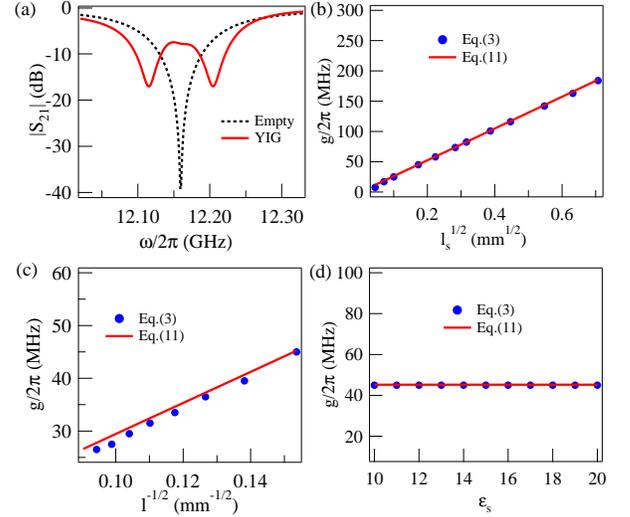,width=8cm}
\caption{(Color online) (a) Simulated $S_{21}$ spectra of $h-$mode for the YIG loaded cavity at $\mu_0H_{ext}$=380 mT (red solid lines) as compared to the cavity mode for the empty cavity (black dotted lines). The coupling strength $g$ as a function of (b) $l_s^{1/2}$ at $W=85$ mm and $\varepsilon_s=15$, (c) $l^{-1\!/2}$ at $l\!_s=0.03$ mm  and $\varepsilon\!_s=15$ and (d) $\varepsilon\!_s$ at $W=85$ mm and $l_s=0.03$ mm. Solid lines are the analytical results based on Eq. (\ref{Eq:coupled h-mode}) and solid circles are extracted from the full-wave simulation based on Eq. (\ref{Eq:S21}).}
\label{Fig5}
\end{figure}

To validate the analytical expression Eq. (\ref{Eq:homdecouplingstrength}) for the $h$-mode, we compare it with the results using full-wave simulation based on Eq. (\ref{Eq:S21}). Parameters for our YIG sample are gyromagnetic ratio of $\gamma=$ 169 $\mu_0$ GHz/T, the saturation magnetization of $\mu_0M_0=$0.155 T and damping factor $\alpha=1.25\times10^{-3}$ obtained from independent measurements, and $\varepsilon_s=15$ is from Ref. \onlinecite{YIG1, YIG2}. As shown in Fig. \ref{Fig5}(a), the coupling of the cavity and FMR modes results in a CMP gap  (=$2g$) at $\omega_{C\!R}=\omega_{F\!M\!R}$, where the magnon-like CMP and the photon-like CMP have an equal amplitude in $S_{21}$ spectra.

For a systematic comparison, the full-wave simulation is performed as a function of $l\!_s$, $l$ and $\varepsilon\!_s$ at the conditions of $\omega_{C\!R}=\omega_{F\!M\!R}$=12.16 GHz (corresponding to $\mu_0H_{ext}$=380 mT) and the resultant coupling strength $g$ is plotted as solid symbols in Figs. \ref{Fig5}(b-d), respectively, where $W=2l+l_s$=85 mm and $\varepsilon\!_s$=15 in Fig. \ref{Fig5}(b), $l\!_s$=0.03 mm and $\varepsilon_s$=15 in Fig. \ref{Fig5}(c), and  $l\!_s$=0.03 mm and $W=85$ mm in Fig. \ref{Fig5}(d). In order to satisfy the condition of $\omega_{C\!R}$=12.16 GHz the odd integer $q$ and $\phi_y$ should be adjusted when varying $l$ during the full-wave simulation. However, both should not affect the calculated coupling strength. The coupling strength $g$ calculated from the analytical expression of Eq. (\ref{Eq:homdecouplingstrength}) is also plotted as a solid line in Fig. \ref{Fig5}(b-d). The agreement between our full-wave simulation and analytical approximation is remarkable. Therefore, Equation (\ref{Eq:homdecouplingstrength}) provides a rigorous relationship between the sample parameters, the geometric parameters of the resonant cavity and the coupling strength, which should benefit not only the study of the CMP but also the design of related microwave devices.

\subsubsection{Coupling strength $g^\prime$ for $e$-modes}

\begin{figure}
\epsfig{file=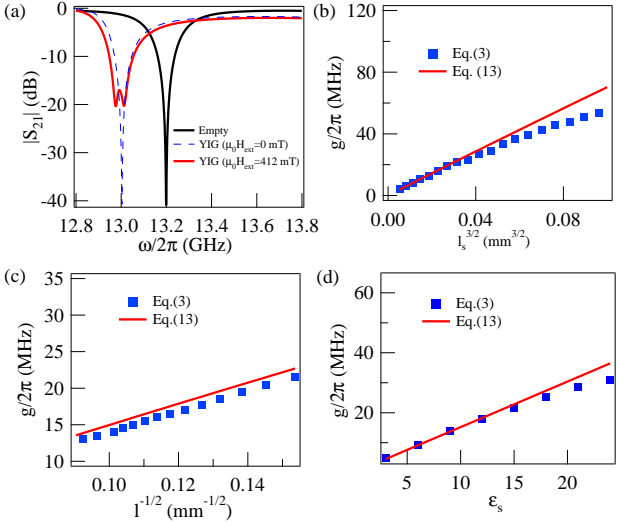,width=8.5cm}
\caption{(Color online) (a) Simulated $S_{21}$ spectra of $e$-mode for empty cavity (black solid lines) and YIG loaded cavity at $H_{ext}$=0 (blue dashed lines). The hybridization of the cavity and FMR modes appears at $\mu_0H_{ext}$=412 mT (red solid lines). The coupling strength $g$ as a function of (b) $l_s^{1/2}$ at $W=85$ mm and $\varepsilon_s=15$, (c) $l^{-1\!/2}$ at $l\!_s=0.1$ mm  and $\varepsilon_s=15$, and (c) $\varepsilon\!_s$ at $W=85$ mm and $l\!_s=0.1$ mm, with full numerical calculation (solid circles) and analytical results (solid lines) using Eq. (\ref{Eq:coupled e-mode}) (red solid lines). }
\label{Fig6}
\end{figure}

For the first order effect, the loaded YIG sample will only result in the red-shift of $e$-modes, which is determined by the electromagnetic property of the sample, expressed as $\delta\omega_{C\!R}=\dfrac{l\!_s}{2l}\omega[(\varepsilon\!_s-1)+(1-\dfrac{1}{\mu\!_s})(\dfrac{\omega\!_c}{\omega_{C\!R}})^2]$ in Eq. (\ref{Eq:e-h mode shift}). This effect is illustrated in Fig. \ref{Fig6}(a), where the full-wave simulation indicates that the cavity resonance for the $e-$mode originally at 13.20 GHz shifts to 13.00 GHz when inserting the YIG sample in the absence of the external magnetic field. More interestingly, the full-wave simulation clearly reveals the resultant CMP gap due to the hybridization of the cavity modes and FMR modes at $H_{ext}=412$ mT (corresponding to $\omega_{F\!M\!R}=\omega_{C\!R}-\delta\omega_{C\!R}$). To analytically explain this effect, the effect due to higher order $l_s$ terms should be taken into account. Different from the analytical approximation for $h$-modes, one should expand Eq. (\ref{Eq:S21}) according to both $\omega-\omega_{C\!R}$ and $l\!_s$, because in this case $\omega-\omega_{C\!R}\approx\delta\omega_{C\!R}$, i.e., $\propto{}l\!_s$. Expanding up to third order in both $\omega-\omega_{C\!R}$ and $l\!_s$ and replacing higher order $\omega-\omega_{C\!R}$ terms with $\omega-\omega_{C\!R}=\delta\omega_{C\!R}$ based on the first order approximation, the transmission coefficient of the CMP for $e$-modes is found to be
\begin{eqnarray}
S_{21}&=&\dfrac{2\sqrt{2}}{3}e^{-i\phi_y} \nonumber\\
&&\times\left[ 1 + \dfrac{\Delta\omega^\prime_{C\!R}}{i(\omega-\omega^\prime_{C\!R})-\Delta\omega^\prime_{C\!R}-\frac{g^\prime{}^2}{i(\omega-\omega_{F\!M\!R})-\Delta\omega_{F\!M\!R}}}\right] \nonumber\\
\label{Eq:coupled e-mode}
\end{eqnarray}

\noindent with
\begin{eqnarray}
&&\omega^\prime_{C\!R}=\omega_{C\!R}-\dfrac{l\!_s}{2l}\omega[\varepsilon\!_s-1+(\omega\!_c/\omega_{C\!R})^2],
\nonumber\\
&&\Delta\omega^\prime_{C\!R}=\Delta\omega_{C\!R}=(c/6l)\sqrt{1-(\omega\!_c/\omega_{C\!R})^2},\nonumber \\
&&g^\prime{}^2=\varepsilon_s^2(\omega _H +\omega _M)\omega_M\dfrac{\omega^2}{c^2}\dfrac{l_s^3}{48l}.
\label{Eq:gforemode}
\end{eqnarray}

Similar to that for the $h-$mode, the coupling strength $g^\prime$ in Eq. (\ref{Eq:gforemode}) is verified by comparing its dependence on $l_s$, $l$ and $\varepsilon_s$ with the results from the full-wave simulation as shown in Fig. \ref{Fig6}(b-d), respectively, where $W=$85 mm and $\varepsilon_s$=15 in Fig. \ref{Fig6}(b),  $l_s$=0.1 mm and $\varepsilon_s$=15  for Fig. \ref{Fig6}(c), and $l\!_s=0.1$ mm and $W=85$ mm for Fig. \ref{Fig6}(d). Without any adjustable parameters, the agreement between the full-wave simulation and the analytical approximation is good. Although $g^\prime\propto{}l^{-1\!/2}$ for the $e-$mode is the same as $g\propto{}l^{-1\!/2}$ for the $h-$ mode, $g^\prime$ is linearly proportional to $l_s^{3\!/2}$  and $\varepsilon_s$, differing from $g$.

\section{Experimental results and discussion}

\begin{figure} [t]
\begin{center}
\epsfig{file=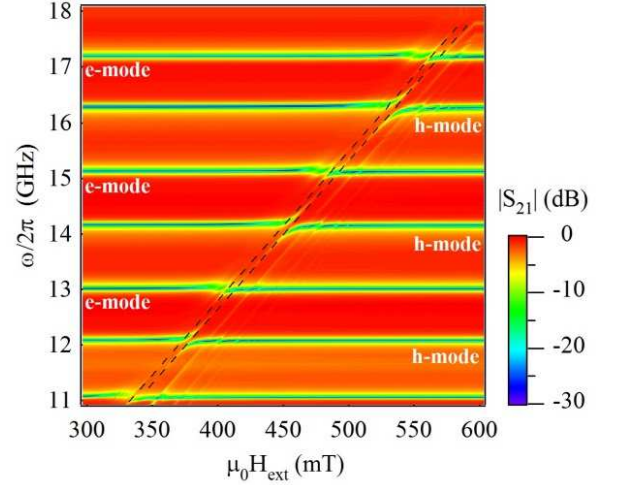,width=8cm}
\caption{(Color online)  $S_{21}$ spectra measured as a function of external magnetic field and microwave frequency for a YIG ($d$=0.5 mm) loaded 1D cavity. Dashed lines show the two strongest coupled magnon modes: the right one is coupled with $h$-modes and the left one with $e$-modes.}
\label{Fig7}
\end{center}
\end{figure}

The CMP experiments have been performed by inserting an YIG sample into the proposed 1D microwave cavity. Due to the difficulty in obtaining YIG samples with an exact diameter of 1.61 cm to satisfy the requirement of our 1D model, a set of rectangular YIG samples with a dimension of about 5.5$\times$3.5 mm$^2$ was used instead. In the first experiment, the YIG sample (with a thickness of $d$=0.5 mm) was inserted into the cavity with a total length of $W=2l+l_s=85$ mm and $S_{21}$ spectra were measured by successively changing the magnetic bias $H_{ext}$ from 300 mT to 600 mT with a step size of 0.1 mT. Figure \ref{Fig7} maps the amplitude of $S_{21}$ as a function of the frequency and the magnetic bias. It has been found that a number of magnon modes in the YIG sample couple to the cavity. The multiple resonance structure could be attributed to the magnetostatic modes\cite{White1956, Dillon1958} in our YIG sample, which have also been observed in previous coupled magnon-photon systems\cite{Goryachev2014, Haigh2015}. A carefully inspection indicates that there are two magnon modes (guided by the dashed lines) strongly coupled with the cavity modes: one coupled with $h$-mode and another coupled with $e$-mode. The separation between them is only 6.0 mT. 

To simplify the complicated 3D wave propagation problem caused by (1) the lateral size ($\sim$0.2 cm$^2$) of the YIG sample, which is much smaller than the cross section ($\sim$2 cm$^2$) of the circular waveguide, and (2) spatially non-uniform precession of the magnon system, a filling factor $\xi=l_s/d$ that converts the thickness $d$ of the YIG sample to be an effective $l_s$ in our 1D model was introduced. As verified in the following experiments, it doesn't affect the dependence of the coupling strength on either $l$ or $l_s$ and furthermore it can quantitatively explain the experimental observation.

\begin{figure} [t]
\begin{center}
\epsfig{file=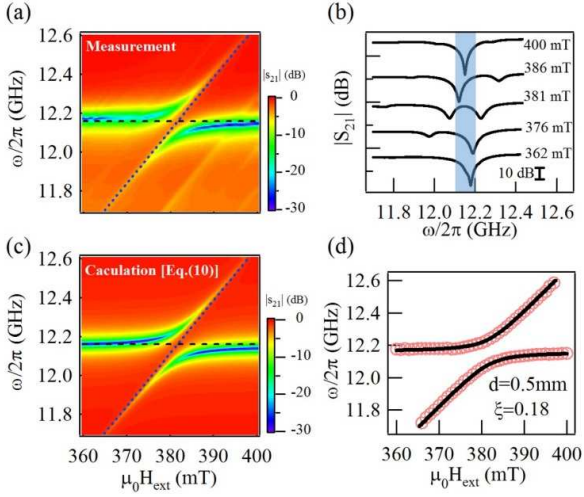,width=8cm}
\caption{(Color online) (a) $S_{21}$ spectra for $h$-mode at 12.157 GHz measured as a function of external magnetic field and microwave frequency. The dotted line refers to the predicted FMR dispersion according to Kittel's formula, while the dashed line is the cavity mode. (b) Typical $S_{21}$ spectra for the CMP. (c) The calculated evolution of hybridized FMR and cavity mode corresponded to $g/2\pi=$ 78 MHz with $\xi$=0.18. (d) Dispersion of the hybridized CMP for measurements (symbols) and calculation (solid line) based on Eq. (\ref{Eq:coupled h-mode}).}
\label{Fig8}
\end{center}
\end{figure}

The coupling of the magnon-photon system near the $h$-mode at $\omega_{C\!R}/2\pi=12.157$ GHz as shown in Fig. \ref{Fig8}(a). When the magnon modes are far from the cavity modes, they follow the dotted line, which is the dispersion calculated from Kittel$'$s formula. However, when the predicated magnon frequency matches with the cavity resonance frequency indicated by the dashed line, the coupling between microwave and magnetization dynamic hybridizes the magnon and cavity modes and a pronounced splitting of cavity modes is observed.

Typical $S_{21}$ spectra in Fig. \ref{Fig8}(b) illustrates the evolution of the hybrid magnon and cavity mode. When the resonance frequency difference between the magnon mode and the cavity mode is large, the magnon is only weakly excited by the electromagnetic field, while the amplitude of the cavity mode is orders of magnitude larger. The amplitude of magnon modes is significantly enhanced when it approaches the cavity mode. This effect is due to the fact that the coupling of magnon and cavity modes generates hybridized cavity photon-magnon quasi-particles, i.e., cavity magnon polaritons\cite{Bai2015}. At $\mu_0H_{ext}=381$ mT corresponding to $\omega_{F\!M\!R}=\omega_{C\!R}$, the photon-like and magnon-like CMP have equal amplitude and occur at $\omega_{C\!R}\pm g$, indicating that the microwave energy is effectively transferred from the photon system to the magnon system through the mutual electrodynamic coupling between them. Similar to the optical system\cite{EIT2005} and metamaterial system\cite{Papasimakis2009}, the coupling between magnon and photon could experience electromagnetic induced transparency and other coupling phenomena\cite{Zhang2014}. 

For comparison, we numerically calculate the $|S_{21}|$ coefficient using Eq. (\ref{Eq:coupled h-mode}) and plot it in Fig. \ref{Fig8}(c), which reproduces the measured coupling feature. The resonance frequencies of the hybrid modes are plotted as a function of the magnetic bias $H_{ext}$ in Fig.\ref{Fig8}(d), where the calculated dispersion (solid line) agrees very well with the measured data (symbols). To match $g/2\pi=$78 MHz deduced from the measured separation between the the hybrid modes at $\mu_0H_{ext}=$381 mT, a filling factor of $\xi$=0.18 is introduced to simplify the complicated 3D problem while other parameters are determined from independent experiments. 

\begin{figure} [t]
\begin{center}
\epsfig{file=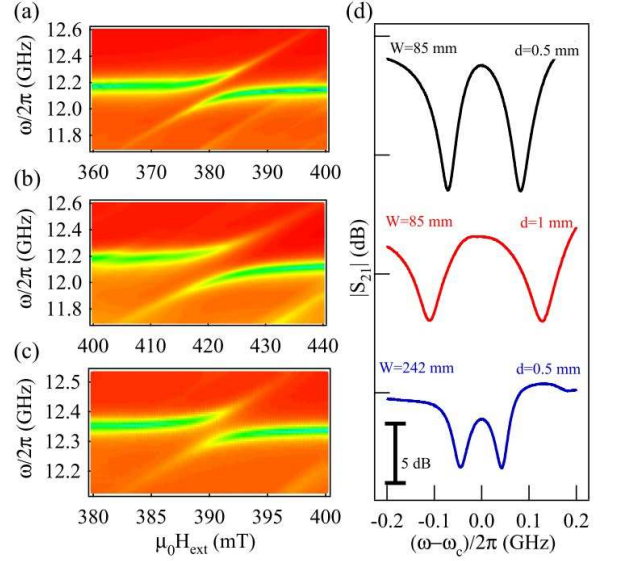,width=8cm}
\caption{(Color online) (a) For $h-$mode $S_{21}$ amplitude mapping as a function of the microwave frequency $\omega$ and the magnetic bias $H_{ext}$ for three different cases: (a)$d=0.5$ mm and $W=85$ mm,  (b) $d=1$ mm and $W=85$ mm, and (c) $d=0.5$ mm and $W=242$ mm. (d) Typical spectra of $S_{21}$ shows the splitting of the photon-like and magnon-like CMP when they have equal amplitude. The deduced coupling strength are 78 MHz, 121 MHz, and 48 MHz. }
\label{Fig9}
\end{center}
\end{figure}

To verify the scaling effect of $l_s$ and $l$ on the coupling strength $g$ for the $h$-mode, the second experiment has been performed near 12.157 GHz by varying $l_s$ and $l$.  Fig. \ref{Fig9}(b) presents the amplitude mapping of $S_{21}$ as a function of the frequency and the magnetic bias in the vicinity of $h$-modes at 12.157 GHz, where a $d$=1 mm thick YIG is loaded while keeping the cavity length $W=85$ mm unchanged. Similar dispersion of the CMP is observed, however, with a much larger separation compared with the dispersion measured for the YIG sample with $d$=0.5 mm. The enhancement of the coupling strength can be clearly revealed by inspecting the $S_{21}$ spectra at $\omega_{F\!M\!R}=\omega_{C\!R}$ as shown in Fig. \ref{Fig8}(b), where $g/2\pi$=121 MHz is deduced. When compared with the value of $g/2\pi$= 78 MHz measured for a 0.5 mm thick sample, the obtained scaling factor is 1.5 which is close to the expected scaling factor of $\sqrt{2}$. 

We also checked the $l$ dependence by elongating the length of the circular waveguide to $W=$242 mm but keeping $d$=0.5 mm unchanged. As expected a weaker coupling was found as shown in Fig. \ref{Fig9}(c). The measured $g/2\pi$=44 MHz is close to the expected value of $g/2\pi$=48 MHz scaled by $l^{-1/2}$. Therefore, the scaling effect of the sample thickness and the length of the cavity is demonstrated for $h$-modes by this experiment. We note that in this experiment the resonance frequency of the magnon mode slightly shifts with the sample thickness and the cavity also moves with $l$. However, a careful inspection indicates that the resultant changes in the coupling strength should be less than 5\%.  

\begin{figure} [t]
\begin{center}
\epsfig{file=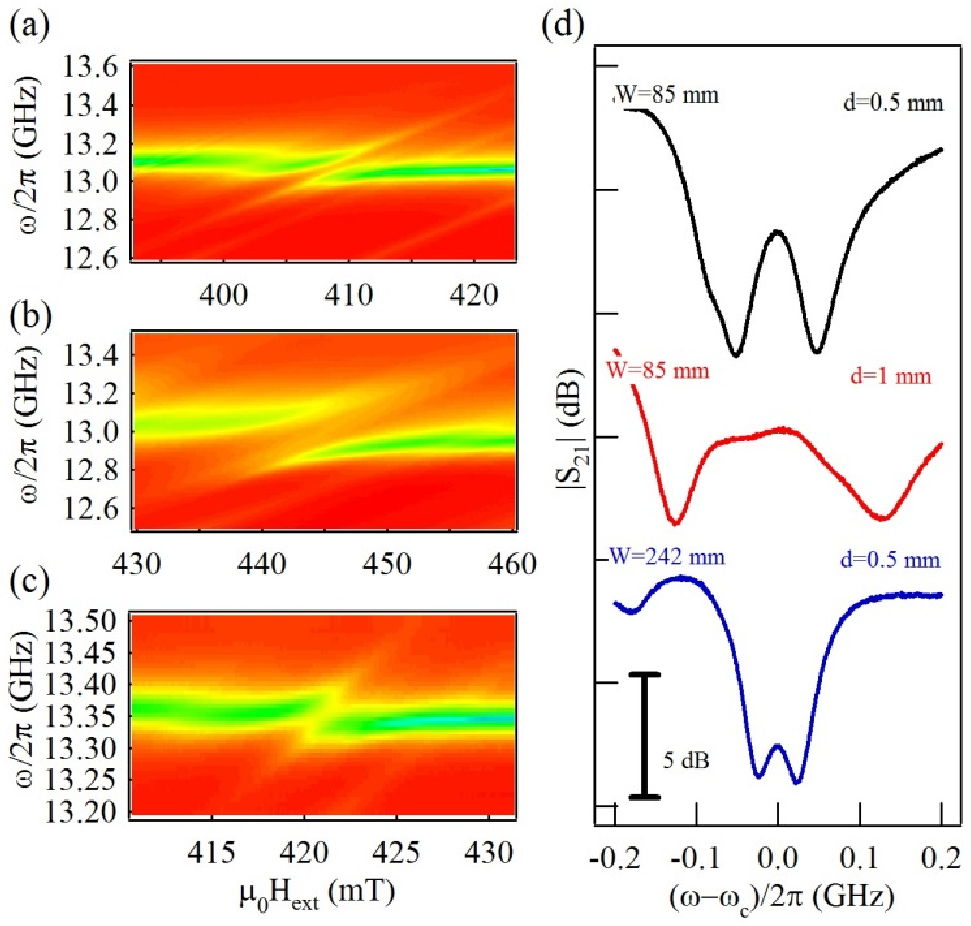,width=8cm}
\caption{(Color online)(a) For $e-$mode $S_{21}$ amplitude mapping as a function of the microwave frequency $\omega$ and the magnetic bias $H_{ext}$ for three different cases: (a) $d=0.5$ mm and $W=85$ mm,  (b) $d=1$ mm and $W=85$ mm, and (c) $d=0.5$ mm and $W=242$ mm. (d) Typical spectra of $S_{21}$ shows the splitting of the photon-like and magnon-like CMP when they have equal amplitude. The deduced coupling strengths $g^\prime$ are 49 MHz, 130 MHz, and 24 MHz.}
\label{Fig10}
\end{center}
\end{figure}

In the third experiment, the scaling effect of $l_s$ and $l$ on the coupling strength $g^\prime$ for the $e$-mode was verified. For the measurement performed for $d=0.5$ mm and $W=85$ mm, the coupling strength is about 49 MHz for the $e-$mode at 13.20 GHz, which is weaker than that of 78 MHz for the $h-$mode at 12.157 GHz. Increasing $d$ to 1 mm the coupling strength $g^\prime$ is raised to 130 MHz as shown in Fig. \ref{Fig10}(b). The estimated scaling factor of $l_s^{1.4}$ for $e-$mode is in agreement with  $l_s^{3/2}$ predicted by our theory Eq. (\ref{Eq:gforemode}). Similarly to the $h$-modes, the coupling strength $g^\prime$ scales approximately with $l^{-1\!/2}$ using the values of 49 MHz and 24 MHz for $W=$85 mm and $W=$242 mm, respectively.

\begin{figure} [t]
\begin{center}
\epsfig{file=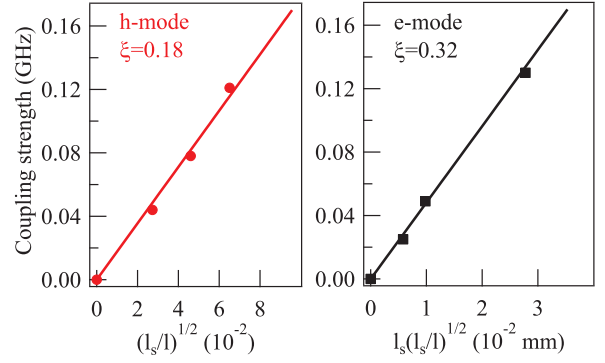,width=8cm}
\caption{(Color online) Coupling strength as a function of $\sqrt{l_s\!/l}$ and $l_s\!\sqrt{l_s\!/l}$ for $h$-mode and $e$-mode, respectively. Symbols are experimental data with $l_s=\xi{d}$ and solid lines are calculation based on the analytic solution using $\mu_0H_{ext}=380$ mT, $\omega_{C\!R}/2\pi=12.2$ GHz for the $h$-mode and $\mu_0H_{ext}=420$ mT, $\omega_{CR}/2\pi=13.2$ GHz for the $e$-mode, respectively.}
\label{Fig11}
\end{center}
\end{figure}

To summarize our experimental observation and theoretical expectation, Figure \ref{Fig11} highlights different dependence of the coupling strength, where $g$ is linearly proportional to $\sqrt{l_s/l}$ while $g^\prime$ is linearly proportional to $l_s\sqrt{l_s/l}$. To fit the analytic results (solid lines) calculated based on Eq. (\ref{Eq:homdecouplingstrength}) and Eq. (\ref{Eq:gforemode}) for the $h$- and $e$-modes, respectively, the filling factors for experimental data (symbols) are found to be $\xi=$0.18 for the $h$-mode occurring at $\sim12.2$ GHz and $\xi=$0.32 for the $e$-mode occurring at $\sim13.2$ GHz. The difference in $\xi$ is believed to be due to the detailed wave propagation of the $h$-mode and $e$-mode. 

\section{Conclusions}
In conclusion, we have developed a transfer matrix approach to investigate the dispersion of the cavity-magnon-polariton (CMP), created by inserting a low damping magnetic insulator into a high-quality 1D microwave cavity. In the limit of long wavelength an analytical approximation of the CMP dispersion has been obtained, where the coupling strength characterizing the interplay between the magnetization and microwave dynamics is determined by the sample thickness, the permittivity, the cavity length as well the the parity of the cavity modes. These scaling effects of the cavity and material parameters is then confirmed by experimental data. The analytic solution indicates the universality of the 1D CMP, in which the coupling strength is not sensitive to the detailed design of the cavity. This work not only provides a detailed physical understanding of the coupled dynamic system that could be used to quantify the characteristic phenomena associated with different coupling regimes, but also could be potentially used for designing dynamic filters and switch devices for microwave applications.

\section{ACKNOWLEDGEMENTS}
This work has been funded by NSERC, CFI, the National Key Basic Research Program of China (2011CB925604), and the National Natural Science Foundation of China Grant No. 11429401. We would like to thank Sandeep Kaur, Lihui Bai, Michael Harder, Hans Huebl, and Yunshan Cao for useful discussions.

\clearpage

\appendix
\section{Details of the wave propagation in the cavity}
\label{Sec:appendixA}
In this Appendix, we present some details of the analytical solution for the wave propagation in the microwave resonance cavity, which can be used for the quantitative analysis of the coupling between FMR and cavity resonance. We start with the transfer matrix (the ABCD matrix in microwave terminology) following the $e^{-i\omega{t}}$ convention. For a cascaded system of many components such as in Fig.\ref{Fig1}(b), the transfer matrix between ports A and B is determined by multiplying the ABCD matrix for each component as \cite{Pozar2012}
\begin{equation}
M=\left(\begin{matrix} A & B\\ C & D \end{matrix}\right)= M_1M\!_sM_2.
\label{SEq:Mmatrix}
\end{equation}

The transfer matrix ($M_1$) of the air-filled waveguide section between port A and port L is
\begin{equation}
M_1=\left(
\begin{matrix}
\cos(k_0{}l) & -iZ_0\sin(k_0{}l)\\
-iZ_0^{-1}\sin(k_0{}l) & \cos(k_0{}l)
\end{matrix}\right),
\end{equation}

\noindent the transfer matrix ($M_s$) for the sample-loaded waveguide (yellow area between port L and R in Fig.\ref{Fig1}(b)) is
\begin{equation}
M\!_s=\left(\begin{matrix}
\cos({k\!_s}{l\!_s}) & -iZ\!_s\sin({k\!_s}{l\!_s})\\
-iZ\!_s^{-1}\sin({k\!_s}{l\!_s}) & \cos({k\!_s}{l\!_s})
\end{matrix}\right),
\end{equation}

\noindent and the transfer matrix ($M_2$) of the air-filled waveguide section between port R and port B is
\begin{equation}
M_2=\left(
\begin{matrix}
\cos(k_0{}l) & -iZ_0\sin(k_0{}l)\\
-iZ_0^{-1}\sin(k_0{}l) & \cos(k_0{}l)
\end{matrix}\right).
\end{equation}

In the above equations the wave number and the characteristic impedance are defined as
\begin{equation}
k\!_s = \dfrac{\omega}{c}\sqrt{\varepsilon\!_s\mu\!_s-\left(\dfrac{\omega\!_c}{\omega}\right)^2}, \mbox{\space{}and\space} {Z\!_s} = \dfrac{\omega {\mu _0}{\mu\!_s}}{k\!_s}.
\label{Eq:appendix kandZ}
\end{equation}

Here $\omega$, $\omega\!_c$, $\varepsilon\!_s$($\mu\!_s$), and $\varepsilon_0$($\mu_0$) are the frequency of the operating microwave, the cut-off frequency of the circular waveguide, the relative complex permittivity (permeability), and vacuum permittivity/permeability, respectively. $c=1/\sqrt{\varepsilon_0\mu_0}$ is the vacuum speed of light. For empty waveguide (air-filled waveguide), the wave number and characteristic impedance are $k_0$ and $Z_0$ using $\varepsilon\!_s=\mu\!_s=1$.

The scattering equations between the ports A and B are expressed as
\begin{equation}
\left(\begin{matrix}e_{A}^{-} \\ e_{B}^{-} \end{matrix}\right)=
\left(\begin{matrix} S_{AA} &S_{AB}\\ S_{BA} &S_{BB}\end{matrix}\right)
\left(\begin{matrix} e_{A}^{+} \\e_{B}^{+}\end{matrix}\right),
\label{SEq:scatteringEquation}
\end{equation}
\noindent where the where the superscript \textquotedblleft$-$\textquotedblright and \textquotedblleft$+$\textquotedblright indicate that the direction of wave propagation is from a circular waveguide to the transition and vice versa.
Thus the S-parameter between port A and B can be calculated as\cite{Pozar2012}
\begin{widetext}
\begin{equation}
\begin{split}
\begin{pmatrix}
S_{AA} & S_{AB}\\ S_{BA} & S_{BB}\end{pmatrix}&=\begin{pmatrix}
{\frac{{A + B/{Z_0} - C{Z_0} - D}}{{A + B/{Z_0} + C{Z_0} + D}}}&{\frac{{2(AD - BC)}}{{A + B/{Z_0} + C{Z_0} + D}}}\\
{\frac{2}{{A + B/{Z_0} + C{Z_0} + D}}}&{\frac{{ - A + B/{Z_0} - C{Z_0} + D}}{{A + B/{Z_0} + C{Z_0} + D}}}
\end{pmatrix}\\
 &=e^{2ik_0l}\begin{pmatrix}
{{{\sin ({k\!_s}{l\!_s})(Z_s^2 - Z_0^2)} \over {(Z_0^2 + Z_s^2)\sin ({k\!_s}{l_s}) + 2i{Z_0}{Z\!_s}\cos ({k\!_s}{l\!_s})}}} &{{{2i{Z_0}{Z\!_s}} \over {(Z_0^2 + Z_s^2)\sin ({k\!_s}{l\!_s}) + 2i{Z_0}{Z_s}\cos ({k\!_s}{l\!_s})}}} \\
{{{2i{Z_0}{Z\!_s}} \over {(Z_0^2 + Z_s^2)\sin ({k\!_s}{l\!_s}) + 2i{Z_0}{Z\!_s}\cos ({k\!_s}{l\!_s})}}} & {{{\sin ({k\!_s}{l\!_s})(Z_s^2 - Z_0^2)} \over {(Z_0^2 + Z\!_s^2)\sin ({k\!_s}{l\!_s}) + 2i{Z_0}{Z\!_s}\cos ({k\!_s}{l\!_s})}}}
\end{pmatrix}
\end{split}
\label{SEq:SAB_matrix}
\end{equation}
\end{widetext}

Notice that the co-polarized TE11 mode exists in the circular waveguide and the mono-polarized TE10 mode exists in the rectangular waveguide.  The resulting boundary conditions for port A and B (in $x-y$ coordinates) are

\begin{subequations}
\begin{eqnarray}
\left(\begin{matrix} e_{Ax}^{+}\\e_{Ay}^{+} \end{matrix}\right)
&=\Gamma _{AA}\left(\begin{matrix} e_{Ax}^{-}\\e_{Ay}^{-} \end{matrix}\right)+
\left(\begin{matrix} e_{Ax}^{inc}\\e_{Ay}^{inc} \end{matrix}\right),\\ 
\left(\begin{matrix} e_{Bx}^{+}\\e_{By}^{+} \end{matrix}\right)
&=\Gamma _{BB}\left(\begin{matrix} e_{Bx}^{-}\\e_{By}^{-} \end{matrix}\right)+
\left(\begin{matrix} e_{Bx}^{inc}\\e_{By}^{inc} \end{matrix}\right),
\label{SEq:Boundary}
\end{eqnarray}
\end{subequations}

\noindent where $e_{Ax,Ay}^{inc}$ and $e_{Bx,By}^{inc}$ are the incident waves from port A and B with $x$ and $y$ polarization, respectively. In the apparatus shown in Fig.\ref{Fig1} $y-$ and $y^\prime-$ polarized TE11 waves entering the transitions from port A and B, are reflected with a reflectivity $R$ and phase jump of $\phi_{y}$, while $x-$ and $x^\prime-$ polarized TE11 waves are transmitted through the transition without reflection. Thus the reflecting matrix $\Gamma_{AA}$ and $\Gamma_{BB}$ can be expressed as
\begin{subequations}
\begin{eqnarray}
\Gamma _{AA}&=&R\left(\begin{matrix} 0&0\\
0&{{e^{i{\phi _y}}}} \end{matrix}\right), \\
\Gamma _{BB}&=&\left(\begin{matrix} \cos(\theta)&-\sin(\theta)\\
\sin(\theta)&\cos(\theta)\end{matrix}\right)
\Gamma _{AA} \left(\begin{matrix} \cos(\theta)&\sin(\theta)\\
-\sin(\theta)&\cos(\theta)\end{matrix}\right). \nonumber \\
\label{Gamma_AAand_BB}
\end{eqnarray}
\end{subequations}

Combining the scattering equations Eqs. (\ref{SEq:scatteringEquation}) between port A and B and using the boundary conditions Eqs. (\ref{SEq:Boundary}) at port A and B, it is found \cite{Sounas2013}
\begin{equation}
\left(\begin{matrix} {1 - {S_{AA}}{\Gamma _{AA}}}&{ - {S_{AB}}{\Gamma _{BB}}}\\
{ - {S_{BA}}{\Gamma _{AA}}}&{1 - {S_{BB}}{\Gamma _{BB}}} \end{matrix}\right)
\left(\begin{matrix} {e_A^ - }\\
{e_B^ - } \end{matrix}\right)
=\left(\begin{matrix} {{S_{AA}}}&{{S_{AB}}}\\
{{S_{BA}}}&{{S_{BB}}} \end{matrix}\right)
\left(\begin{matrix} {e_A^{inc}}\\
{e_B^{inc}} \end{matrix}\right)
\label{SEq:relation between E and Einc}
\end{equation}

Taking into account the $x-y$ co-polarized TE11 wave, Eq. (\ref{SEq:relation between E and Einc}) describes the relation between the transmitting/reflecting wave and incident wave in the $x-y$ coordinate, which can be written as
\begin{widetext}
\begin{equation}
\begin{matrix}
\left( \begin{matrix}
e_{Ax}^ - \\
e_{Ay}^ - \\
e_{Bx}^ - \\
e_{By}^ -
\end{matrix} \right) = {\left( {\begin{matrix}
{\left( {\begin{matrix}
1&0\\
0&1
\end{matrix}} \right) - \left( {\begin{matrix}
{{S_{AA}}}&0\\
0&{{S_{AA}}}
\end{matrix}} \right){\Gamma _{AA}}}&{ - \left( {\begin{matrix}
{{S_{AB}}}&0\\
0&{{S_{AB}}}
\end{matrix}} \right){\Gamma _{BB}}}\\
{ - \left( {\begin{matrix}
{{S_{BA}}}&0\\
0&{{S_{BA}}}
\end{matrix}} \right){\Gamma _{AA}}}&{\left( {\begin{matrix}
1&0\\
0&1
\end{matrix}} \right) - \left( {\begin{matrix}
{{S_{BB}}}&0\\
0&{{S_{BB}}}
\end{matrix}} \right){\Gamma _{BB}}}
\end{matrix}} \right)}
{\left( {\begin{matrix}
{{S_{AA}}}&0&{{S_{AB}}}&0\\
0&{{S_{AA}}}&0&{{S_{AB}}}\\
{{S_{BA}}}&0&{{S_{BB}}}&0\\
0&{{S_{BA}}}&0&{{S_{BB}}}
\end{matrix}} \right)^{ - 1}}\left( {\begin{matrix}
{e_{Ax}^{inc}}\\
{e_{Ay}^{inc}}\\
{e_{Bx}^{inc}}\\
{e_{By}^{inc}}
\end{matrix}} \right)
\end{matrix}
\label{SEq:fullmatrix}
\end{equation}
\end{widetext}

As the experimentally accessible parameters are the transmission coefficients $S_{12}$ and/or $S_{21}$ between ports 1 and 2 of the assembled waveguide cavity, defined as
\begin{subequations}
\begin{eqnarray}
{S_{21}} &=& \frac{{e_{Bx'}^ - }}{{e_A^{inc}}} = \frac{{e_{Bx}^ - \cos (\theta ) - e_{By}^ - \sin (\theta )}}{{e_A^{inc}}}, \\
 {S_{12}} &=& \frac{{e_{Ax}^ - }}{{e_B^{inc}}}.
\end{eqnarray}
\end{subequations}

By solving Eq. (\ref{SEq:fullmatrix}) the final transmission coefficients can be obtained:
\begin{widetext}
\begin{equation}
 S_{12}=S_{21}=\frac{{\cos (\theta ){S_{AB}}[ - 1 + {R^2}{e^{2i{\phi _y}}}{S_{AB}}{S_{BA}} + {e^{i{\phi _y}}}R{S_{BB}} - {e^{i{\phi _y}}}R{S_{AA}}( - 1 + {e^{i{\phi _y}}}R{S_{BB}})]}}{{ - 1 + {e^{2i{\phi _y}}}{{\cos }^2}(\theta ){R^2}{S_{AB}}{S_{BA}} + {e^{i{\phi _y}}}R{S_{BB}} - {e^{i{\phi _y}}}R{S_{AA}}( - 1 + {e^{i{\phi _y}}}R{S_{BB}})}}
 \label{SEq:S21andS12}
\end{equation}
\end{widetext}

\section{Wave propagation in the vicinity of the ferromagnetic resonance}
\label{Sec:appendixB}
In this Appendix, we present some details of the analytical solution for the wave propagation near the FMR. In the vicinity of the FMR frequency, the wave propagation is related to the dipolar field associated with the interaction of spins from Maxwell's equations, written as

\begin{subequations}
\begin{eqnarray}
\nabla\times\mathbf{h}&=\varepsilon_s\varepsilon_0\dfrac{\partial\mathbf{e}}{\partial{t}}+\sigma\mathbf{e},
\label{SEq:Maxwell's equation1} \\
\nabla\times\mathbf{e}&=-\mu_0\dfrac{\partial{(\mathbf{h+m}})}{\partial{t}},
\label{SEq:Maxwell's equation2}
\end{eqnarray}
\end{subequations}

\noindent where $\sigma$ is the conductivity of the ferromagnetic sample. For the insulating ferrite YIG $\sigma$ tends to be zero.

Eliminating $\mathbf{e}$ from Eq. (\ref{SEq:Maxwell's equation1}) by taking the curl of it and substituting by Eq. (\ref{SEq:Maxwell's equation2}), we obtain
\begin{equation}
\nabla^2\mathbf{h}-\varepsilon\!_s\varepsilon_0\mu_0\omega^2\mathbf{h}=
\nabla(\nabla\cdot\mathbf{h})+\varepsilon\!_s\varepsilon_0\mu_0\omega^2\mathbf{m}.
\label{SEq:Maxwellsequationsum}
\end{equation}

The motion of $\mathbf{M}$ is governed by the phenomenological Landau-Lifshitz-Gilbert equation\cite{Gilbert2004},
\begin{equation}
\frac{d\mathbf{M}}{dt}=-\gamma(\mathbf{M}\times\textbf{H}_i)+\frac{\alpha}{M_s}
\left(\mathbf{M} \times \frac{d\mathbf{M}}{dt}\right).
\label{SEq:LLG}
\end{equation}

Here $\gamma=\mu_0e/m_e$ is the effective gyromagnetic ratio of an electron (charge $e$ and mass $m_e$), $\alpha$ is the Gilbert damping parameter, $M_s$ is the saturation magnetization. Inside the ferrite sample, the magnetization $\mathbf{M}=\textbf{M}_{0}+\textbf{m}$ and  magnetic field $\mathbf{H}_i=\mathbf{H}_{ext}+\mathbf{h}$ consist of static components $\mathbf{M_0}$ and $\mathbf{H}_{ext}$ as well as the dynamic components $\mathbf{m}$ and $\mathbf{h}$ including a time dependent term of $e^{-i\omega{t}}$. Assuming $H_{ext}$ along $y$ direction, from the vector relation of Eq. (\ref{SEq:LLG}) we can obtain the well-known Polder tensor $\widehat{\chi}$ expressed as
\begin{equation}
\widehat{\chi}=\left(\begin{matrix}
\chi_L &0 &i\chi_T\\
0 &0 &0\\
-i\chi_T &0 &\chi_L
\end{matrix}\right),
\end{equation}
\noindent where

\begin{subequations}
\begin{eqnarray}
\chi_L&=\dfrac{\gamma(\gamma{}H_{ext}-i\alpha\omega)M_0}{(\gamma{}H_{ext}-i\alpha\omega)^2-\omega^2},\\ 
\chi_T&=\dfrac{\gamma{}M_0\omega}{(\gamma{}H_{ext}-i\alpha\omega)^2-\omega^2}.
\label{SEq:Polder tensor}
\end{eqnarray}
\end{subequations}

Using the approximation of long wavelength, i.e., the wave propagates only along the $z$ direction, the coupling equation of Maxwell's equations (\ref{SEq:Maxwellsequationsum}) and Landau- Lifshitz-Gilbert equation (\ref{SEq:LLG}) become
\begin{equation}
\left(\begin{matrix}
\varepsilon\!_s\varepsilon_0\mu_0\omega^2(1+\chi_L)-k_s^2
&i\varepsilon\!_s\varepsilon_0\mu_0\omega^2\chi_T\\
-i\varepsilon\!_s\varepsilon_0\mu_0\omega^2\chi_T
&\varepsilon\!_s\varepsilon_0\mu_0\omega^2(1+\chi_L)
\end{matrix}\right)
\left(\begin{matrix}
h_x\\
h_z
\end{matrix}\right)=0.
\end{equation}

The resultant wave number is
\begin{equation}
k_s^2=\mu_{e\!f\!f}\varepsilon\!_s\varepsilon_0\mu_0\omega^2,
\label{SEq:ksofYIG}
\end{equation}
\noindent where the effective permeability $\mu_{e\!f\!f}$ is defined by $\mu_{e\!f\!f}=[(1+\chi_L)^2-\chi_T^2]/(1+\chi_L)$.


\end{document}